\documentclass[twocolumn,showpacs,preprintnumbers]{revtex4}
\usepackage{mathrsfs}
\usepackage{amssymb}
\usepackage{amsmath}
\usepackage{graphicx}

\begin{document}

\title{Storage of polarization-encoded cluster states in an atomic system}
\author{Chun-Hua Yuan}
\author{Li-Qing Chen}
\author{Weiping Zhang}
\affiliation{State Key Laboratory of Precision Spectroscopy, Department of Physics, East
China Normal University, Shanghai 200062, P. R. China}
\date{\today }

\begin{abstract}
We present a scheme for entanglement macroscopic atomic ensembles
which are four spatially separate regions of an atomic cloud using
cluster-correlated beams. We show that the cluster-type
polarization-encoded entanglement could be mapped onto the
long-lived collective ground state of the atomic ensembles, and the
stored entanglement could be retrieved based on the technique of
electromagnetically induced transparency. We also discuss the
efficiency of, the lifetime of, and some quantitative restrictions
to the proposed quantum memory.
\end{abstract}

\maketitle


\section{Introduction}

Multiparticle graph states, so-called cluster states \cite{cluster},
have attracted much attention for its potential applications as a
basic resource in the ``one-way" quantum computing scheme
\cite{briegel,Nielsen}. The cluster state encoded by the
polarization states of photons has been demonstrated
experimentally \cite%
{clusterexp,Kiesel,Zhang,Prevedel,Lu,Vallone,Chenprl,Tokunaga}.
Meanwhile, the combination of optical techniques with quantum memory
using atoms has shown apparent usefulness to scalable all-optical
quantum computing network \cite{Knill} and long-distance quantum
communication \cite{Duan}. Several experiments in these aspects have
been realized, such as the storage and retrieval of coherent states
\cite{Julsgaard}, single-photon wave packets \cite{Chaneli,Eisaman},
and squeezed vacuum state \cite{Honda,Appel}. With these
achievements, it is worth initiating a study on a reversible memory
for a cluster state.

As we know, light is a natural carrier of classical and quantum information.
Now, the macroscopic atomic system can be efficiently used for its storage
due to long lived electronic ground-state as storage units. Based on the
light and atomic ensembles, two types of quantum memories have been put
forward: one based on the quantum Faraday effect supplemented by measurement
and feedback \cite{Julsgaard}, and the other involving electromagnetically
induced transparency (EIT) \cite{Liu01,Fleischhauer00} and Raman processes
\cite{Kozhekin,Laurat06}. In addition, photon echo technique has also been
proposed for quantum memory \cite{Moiseev}. EIT is probably the most
actively studied technique to achieve a quantum memory. EIT polaritons (dark
state) were first considered theoretically by Mazets and Matisov \cite%
{Mazets}, and later by Fleischhauer and Lukin \cite%
{Fleischhauer00,Fleischhauer2} who suggested storing a probe pulse (stopping
the polariton) by adiabatically switching off the control laser. Extending
the analysis to a double-$\Lambda $ system \cite{Raczy,Li,Chong}, it is
possible to simultaneously propagate and store two optical pulses in the
medium based on a dark state polariton (DSP) consisting of low-lying atomic
excitations and photon states of two frequencies. The existence of the DSP
in the double-$\Lambda $ atomic system studied by Chong \emph{et al.} \cite%
{Chong} required that the fields obey certain conditions for frequency,
amplitude, and phase matchings. Quantitative relations in the case of double-%
$\Lambda $ type atoms are essentially more complicated than for the standard
$\Lambda $ configuration. If one of the conditions breaks, such as the phase
is mismatched, then one of the two pulses will be absorbed and lost \cite%
{Kang}. In this sense, the double-$\Lambda $ system is limited to
the storage of two or more optical pulses. Recently, Yoshikawa
\emph{et al.} \cite{Yoshikawa} demonstrated holographic storage of
two coherence gratings in a single BEC cloud. Choi \emph{et al.}
\cite{Choi} demonstrated that the entanglement of two light fields
survives quantum storage using two atomic ensembles in a cold cloud,
where they realized the coherent conversion of photonic entanglement
into and out of a quantum memory. Hence, in principle, it is now
achievable to store more than two optical fields in atomic ensembles
for different purposes.

In this paper, we propose a scheme that can store a polarization-encoded
cluster state reversibly in a cold atomic cloud based on the EIT technique.
To our best knowledge, the storage of polarization entangled state is very
useful in polarization-encoded quantum computing schemes, such as
\textquotedblleft one-way\textquotedblright\ quantum computing and quantum
coding. On the other hand, our scheme also presents a natural extension of
existing work \cite{Choi,Lukin00}.

Our paper is organized as follows. In Sec.~\ref{polarization}, we describe
how the polarization-encoded cluster state can be stored and retrieved, and
the method of the measurement and verification of entanglement storage. In
Sec.~\ref{analysis}, we analyze and evaluate the efficiency and the
fidelity. In Sec.~\ref{lifetime}, we evaluate the memory lifetime. In Sec.~%
\ref{Dis}, we discuss some restrictions of the proposed quantum memory.
Finally, we conclude with a summary of our results.

\section{Storage of Polarization-encoded Cluster State}

\label{polarization}
\begin{figure}[tbp]
\centerline{\includegraphics[scale=0.38,angle=0]{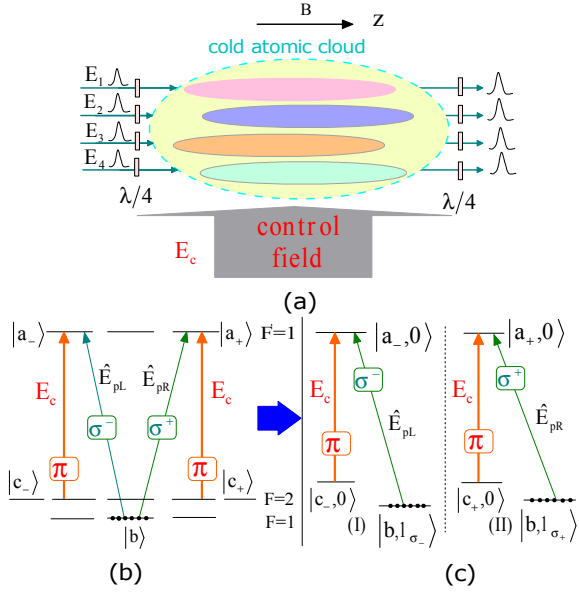}}
\caption{ (Color online) (a) Schematic diagram of the proposed
experiment. The polarization-encoded four-photon cluster state is
inputted with a common perpendicularly-propagating control field
$E_{c}$. Four atomic sub-ensembles (or channels) are represented by
four spatially separate and symmetric regions in a single cloud of
cold atoms. The cold atomic cloud is initially prepared in a
magneto-optical trap (MOT) and the MOT fields are turned off during
the period of the storage and retrieval process. The quantization
axis $z$ is set by the trapping magnetic field $\vec{B}$ in the
preparation of the cold atomic cloud. The horizontally and
vertically polarized single-photon pulses pass through
$\protect\lambda /4$\ plates and are converted into circularly left-
and right-polarized photons, respectively,
i.e., $|H\rangle\rightarrow|\protect\sigma^{-}\rangle$, and $%
|V\rangle\rightarrow|\protect\sigma^{+}\rangle$. (b) The atomic level
configuration for the proposal. $E_{c}$ is the control light with $\protect%
\pi $-polarization, and $\hat{E}_{pL}$ and $\hat{E}_{pR}$ are left and right
circularly-polarized lights, respectively. (c) The corresponding
decompositions of atom-photon couplings in (b) according to the photon
polarizations (I for $\protect\sigma^{-}$ polarization, II for $\protect%
\sigma^{+}$ polarization). }
\label{fig1}
\end{figure}

In this section, we show the EIT technique can be used to realize a
reversible memory for the polarization encoded cluster state \cite{cluster}
\begin{eqnarray}
|\phi _{\text{in}}\rangle &=&\frac{1}{2}[|H\rangle _{1}|H\rangle
_{2}|H\rangle _{3}|H\rangle _{4}+|V\rangle _{1}|V\rangle _{2}|H\rangle
_{3}|H\rangle _{4}  \notag \\
&+&|H\rangle _{1}|H\rangle _{2}|V\rangle _{3}|V\rangle _{4}-|V\rangle
_{1}|V\rangle _{2}|V\rangle _{3}|V\rangle _{4}],
\end{eqnarray}%
where $|H\rangle $ and $|V\rangle $ stand for the single-photon
states with the photon polarized horizontally and vertically,
respectively. The four-photon
cluster state shown above has been demonstrated experimentally \cite%
{clusterexp,Kiesel,Zhang,Prevedel,Lu,Vallone,Chenprl,Tokunaga} using
different methods. Without loss of generality, here we consider the simple
case that the frequencies of all four photons in the state are degenerate.
The schematic diagrams of the proposed experimental system are shown in Fig.~%
\ref{fig1}(a) (hereafter noted case $1$) and Fig.~\ref{fig2}(a) (hereafter
noted case $2$), where four atomic sub-ensembles (or channels) are used.
These four sub-ensembles form four equivalent channels, each of which is
used to store the corresponding polarization-encoded single-photon state $%
(|H\rangle _{i}$ or $|V\rangle _{i}$; $i=1,2,3,4)$ in the four-photon
cluster state.

\subsection{Quantum memory for a polarization-encoded single-photon state}

\begin{figure}[tbph]
\centerline{\includegraphics[scale=0.4,angle=0]{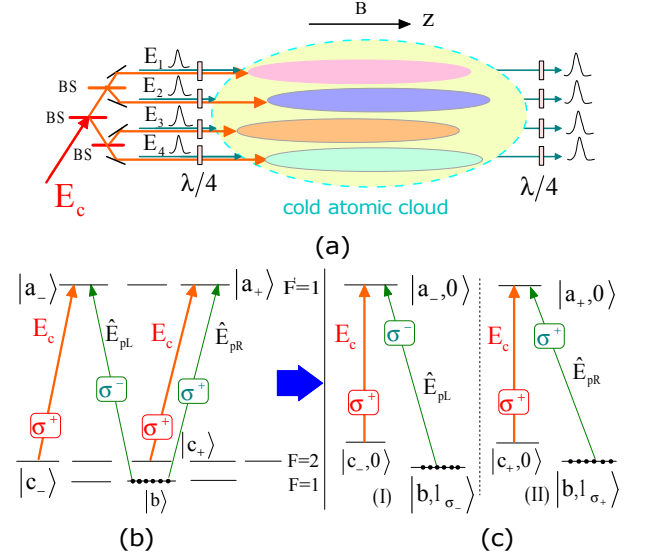}} \caption{
(Color online) (a) The figure is the same as Fig.~1(a) except that a
common collinearly-propagating control field $E_{c}$ is used. (b)
and (c) are
the same as Figs.~1(b) and 1(c), respectively, except the control field $%
E_{c}$ is now replaced with right circularly-polarized light in the
collinearly-propagating case. The right circularly-polarized light
$E_{c}$ is converted by $\lambda/4$ plates using the vertically
polarized light $E_{c}$.} \label{fig2}
\end{figure}

In order to understand the physics behind the schemes shown in Figs. 1 and
2, first we discuss the quantum memory for a polarization-encoded
single-photon state. An atomic ensemble containing $N$ atoms for memory
using the $\Lambda $-type atomic level configuration with the excited state $%
|a\rangle $ and ground states $|b\rangle $ and $|c\rangle $ based on
EIT was studied in detail by Fleischauer and Lukin
\cite{Fleischhauer00} and also reviewed by a few authors, such as
Petrosyan \cite{Petrosyan}. Here, we focus on the case of a
single-photon probe field with horizontal or vertical polarization
and describe the dark state of the system. In the frame rotating
with the probe and the driving field frequencies, the interaction
Hamiltonian is given by
\cite{Fleischhauer00,Fleischhauer2,Petrosyan}
\begin{equation}
\hat{H}=\hbar \sum_{j=1}^{N}[-g\hat{\sigma}_{ab}^{j}\hat{\varepsilon}%
(z_{j})e^{ik_{p}z_{j}}-\Omega _{c}(t)\hat{\sigma}%
_{ac}^{j}e^{ik_{c}^{||}z_{j}}+\text{H.c.}],  \label{single1}
\end{equation}%
where $\hat{\sigma}_{\mu \nu }^{j}=|\mu \rangle _{jj}\langle \nu |$ is the
transition operator of the $j$th atom between states $|\mu \rangle $ and $%
|\nu \rangle $, and we consider the single- and two-photon resonance cases. $%
g=\wp \sqrt{\omega /2\hbar \epsilon _{0}V}$ is the coupling constant between
the atoms and the quantized field mode which for simplicity is assumed to be
equal for all atoms. $k_{p}$ and $k_{c}^{||}=\overrightarrow{k}_{c}\cdot
\overrightarrow{e}_{z}$\ are the wave vectors of the probe field and the
control field along the propagation axis $z$, respectively. The
traveling-wave quantum field operator $\hat{\varepsilon}(z,t)=\sum_{q}\hat{a}%
_{q}(t)e^{iqz}$ is expressed through the superposition of bosonic operators $%
\hat{a}_{q}(t)$ for the longitudinal field modes $q$ with wavevectors $k+q$,
where the quantization bandwidth $\delta q$ $[q\in \{-\delta q/2,\delta
q/2\}]$ is narrow and is restricted by the width of the EIT window $\Delta
\omega _{tr}$ ($\delta q\leq \Delta \omega _{tr}/c$) \cite{Lukin2000}.

Hamiltonian (\ref{single1}) has a family of dark eigen-states $%
|D_{n}^{q}\rangle $\ with zero eigenvalue $\hat{H}|D_{n}^{q}\rangle =0$,
which are decoupled from the rapidly decaying excited state $|a\rangle $.
For a single-photon probe field $n=1$, the dark eigen-states $%
|D_{1}^{q}\rangle $ are given by
\begin{equation}
|D_{1}^{q}\rangle =\cos \theta |1^{q}\rangle |c^{(0)}\rangle -\sin \theta
|0^{q}\rangle |c^{(1)}\rangle ,  \label{dark}
\end{equation}%
where $\theta (t)$ is the mixing angle and $\tan ^{2}\theta
(t)=g^{2}N/|\Omega |^{2}$, and $\theta $ is independent of the mode $q$. $%
|n^{q}\rangle $ denotes the state of the quantum field with $n$
photons in mode $q$, and $|c^{(n)}\rangle $ is a symmetric
Dicke-type state of the
atomic ensemble with $n$ Raman (spin) excitations, i.e., atoms in state $%
|c\rangle $, defined as
\begin{eqnarray}
|c^{(0)}\rangle &=&|b_{1},...,b_{N}\rangle ,~ \\
|c^{(1)}\rangle &=&\frac{1}{\sqrt{N}}%
\sum_{j=1}^{N}e^{i(k+q-k_{c}^{||})z_{j}}|b_{1},b_{2},...,c_{j},...,b_{N}%
\rangle .~~  \label{c1}
\end{eqnarray}

\begin{figure}[tbp]
\centerline{\includegraphics[scale=0.45,angle=0]{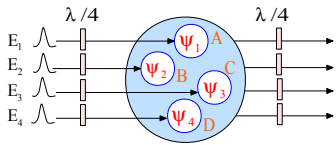}}
\caption{ (Color online) Transverse section of a cold atom cloud.
Once four polarization-entangled single-photon fields have entered
the EIT medium, each
field is converted into corresponding single-excitation polariton $\protect%
\psi _{j}$ $(j=1,2,3,4)$ representing a coupled excitation of the field and
atomic coherence.}
\label{transverse}
\end{figure}

Now, we consider a memory for a single photon with horizontal or vertical
polarization. By the $\lambda /4$\ plate, its polarization is converted into
the left circular or right circular polarization. Due to the propagating
directions of the probe field and the control field, the corresponding
five-level structures are shown in Fig.~\ref{fig1}(b) for case $1$ and~Fig. %
\ref{fig2}(b) for case $2$ instead of the simple $\Lambda $-type level
configuration, where the fields $E_{pL\text{ }}$and $E_{pR}$ interact with
the atoms on the respective transitions $|b\rangle \rightarrow |a_{-}\rangle
$ and $|b\rangle \rightarrow |a_{+}\rangle $, and the excited states $%
|a_{\pm }\rangle $ couple to the metastable states $|c_{\pm }\rangle
$ by the same control field $E_{c}$. According to mapping to the
different metastable states $|c_{+}\rangle $ and $|c_{-}\rangle $,
collective state (\ref{c1}) is written as
\begin{eqnarray}
|c_{+}^{(1)}\rangle &=&\frac{1}{\sqrt{N}}%
\sum_{j=1}^{N}e^{i(k+q-k_{c}^{||})z_{j}}|b_{1},b_{2},...,(c_{+})_{j},...,b_{N}\rangle ,
\notag \\
|c_{-}^{(1)}\rangle &=&\frac{1}{\sqrt{N}}%
\sum_{j=1}^{N}e^{i(k+q-k_{c}^{||})z_{j}}|b_{1},b_{2},...,(c_{-})_{j},...,b_{N}\rangle .
\notag
\end{eqnarray}%
When $\theta =0$ $(|\Omega |^{2}\gg g^{2}N)$, dark state
(\ref{dark}) is comprised of purely photonic excitation, i.e.,
$|D_{1}^{q}\rangle =|1^{q}\rangle |c^{(0)}\rangle $ while in the
opposite limit of $\theta =\pi /2$ $(|\Omega |^{2}\ll g^{2}N)$, it
occurs with the collective atomic
excitation $|D_{1}^{q}\rangle =-|0^{q}\rangle |c_{-}^{(1)}\rangle $ or $%
|D_{1}^{q}\rangle =-|0^{q}\rangle |c_{+}^{(1)}\rangle $ for mapping a left
or a right circularly polarized single-photon. For intermediate values of
mixing angle $0<\theta <\pi /2$, the dark state represents a coherent
superposition of photonic and atomic Raman excitations (polariton) \cite%
{Fleischhauer00,Fleischhauer2}. The left or right circularly polarized
single-photon is converted into respective I class or II class polariton:
\begin{eqnarray}
\text{I class} &\text{:}&\text{ }\hat{\Psi}_{\text{I}}=\cos \theta _{1}(t)%
\hat{\varepsilon}_{\sigma _{-}}-\sin \theta _{1}(t)\sqrt{N}\hat{\sigma}%
_{bc_{-}}, \\
\text{II class} &\text{:}&\text{ }\hat{\Psi}_{\text{II}}=\cos \theta _{2}(t)%
\hat{\varepsilon}_{\sigma _{+}}-\sin \theta _{2}(t)\sqrt{N}\hat{\sigma}%
_{bc_{+}},
\end{eqnarray}%
where $\hat{\sigma}_{bc_{j}}$ $(j=-,+)$ are slowly varying operators which
are defined by $\hat{\sigma}_{\mu \nu }(z,t)=1/N_{z}\sum_{j=1}^{N_{z}}\hat{%
\sigma}_{\mu \nu }^{j}$ with $N_{z}=(N/L)dz\gg 1$.

\subsection{Quantum memory for polarization-encoded four-photon cluster state%
}

In this subsection, we describe the memory for the polarization-encoded
four-photon cluster state. The cold atomic cloud is released from a
magneto-optical trap (MOT), and the quantization axis $z$ is set along the
long axis of the cloud by a small axial magnetic field. The $\pi $-polarized
and the right circularly polarized control fields are respectively used in
case $1$ and case $2$. As a specific example for realization of our scheme
proposed here, we consider hyperfine levels of $^{87}$Rb. For case $1$ [see Fig.~%
\ref{fig1}(b)]: the ground state $|b\rangle $ corresponds to the
$m_{F}=0$ sublevel of the $F=1$, and the states $|c_{-}\rangle $ and
$|c_{+}\rangle $ correspond to the $m_{F}=-1$ and $m_{F}=1$
sublevels of the $F=2$, respectively. The excited states
$|a_{-}\rangle $ and $|a_{+}\rangle $ correspond to the $m_{F}=-1$
and $m_{F}=1$ sublevels of the $F^{\prime }=1$, respectively.
Differently from case $1$, in case $2$ [see Fig.~\ref{fig2}(b)]
because of the control field taking place the $\sigma ^{+}$
transitions, the states
$|c_{-}\rangle $ and $|c_{+}\rangle $ correspond to the $m_{F}=-2$ and $%
m_{F}=0$ sublevels of the $F=2$, respectively.

The outline of our scheme is as follows. All the atoms are initially
prepared in the ground state $|b\rangle $ by optical pumping. The
cold cloud was illuminated by a resonant control laser from radial
and axis directions for cases $1$ and $2$, respectively. The excited
states $|a_{-}\rangle $ and $|a_{+}\rangle $ are resonantly coupled
by the same control field $E_{c}$. Then the polarization-encoded
four-photon cluster state is sent to the four atomic ensembles $A$,
$B$, $C$ and $D$ which are represented by four spatially separate
and symmetric regions in a single cloud of cold atoms
\cite{Choi,Matsukevich,Chou07,Simon,Yoshikawa09}. Before passing
into the EIT medium, the vertically and horizontally polarized
single-photons first pass through $\lambda /4$\ plates and are
converted into circularly right- and left-polarized single-photons,
i.e., $|V\rangle \longrightarrow |\sigma ^{+}\rangle $ and
$|H\rangle \longrightarrow |\sigma ^{-}\rangle $. Once four
polarization-entangled single-photon fields have entered the EIT
medium, these single-photons propagate along the $z$ axis resonantly
interacting with the atoms and making transitions $|b\rangle
\rightarrow |a_{-}\rangle $ or $|b\rangle \rightarrow |a_{+}\rangle
$, and their group velocities are strongly modified by the control
field $E_{c}$. The interaction of the different-polarization
single-photon probe fields with four atomic sub-ensembles separates
into two classes of $\Lambda $-type EIT, which is illustrated in
Figs.~\ref{fig1}(c) and~\ref{fig2}(c). Each probe field is
converted into corresponding single-excitation polariton $\psi _{j}$ $%
(j=1,2,3,4)$ representing a coupled excitation of the field and atomic
coherence, and each polariton $\psi _{j}$ is described by polariton $\hat{%
\Psi}_{\text{I}}$ or $\hat{\Psi}_{\text{II}}$. The corresponding transverse
section of the four atomic sub-ensembles is shown in Fig.~\ref{transverse}.
By switching off the control field adiabatically, these coupled excitations
are converted into the spin wave excitations with a dominant DSP component,
i.e., the cluster state is stored. After a storage period $\tau $, the
stored field state can be retrieved by turning on $E_{c}$ adiabatically.

Once the four single-photons completely enter the EIT medium, under
the adiabatic condition, the state of atomic ensembles $A$, $B$, $C$
and $D$ will adiabatically follow the specific eigen states of the
Hamiltonian (dark states). Then the dark states of the system are
the direct products of those corresponding to the subsystems $A$,
$B$, $C$ and $D$, and the system state vector $|\Phi (t)\rangle $ is
given by
\begin{eqnarray}
\underset{\{q_{1},q_{2},q_{3},q_{4}\}}{|\Phi (t)\rangle } &=&|\mathbf{D}%
_{1}^{q_{1}}\rangle _{A}|\mathbf{D}_{1}^{q_{2}}\rangle _{B}|\mathbf{D}%
_{1}^{q_{3}}\rangle _{C}|\mathbf{D}_{1}^{q_{4}}\rangle _{D}  \notag \\
&=&(\cos \theta _{1}|1^{q_{1}}\rangle |c^{(0)}\rangle _{A}-\sin \theta
_{1}|0^{q_{1}}\rangle |c^{(1)}\rangle _{A})  \notag \\
&\times &(\cos \theta _{2}|1^{q_{2}}\rangle |c^{(0)}\rangle _{B}-\sin \theta
_{2}|0^{q_{2}}\rangle |c^{(1)}\rangle _{B})  \notag \\
&\times &(\cos \theta _{3}|1^{q_{3}}\rangle |c^{(0)}\rangle _{C}-\sin \theta
_{3}|0^{q_{3}}\rangle |c^{(1)}\rangle _{C})  \notag \\
&\times &(\cos \theta _{4}|1^{q_{4}}\rangle |c^{(0)}\rangle _{D}-\sin \theta
_{4}|0^{q_{4}}\rangle |c^{(1)}\rangle _{D})  \notag \\
&=&\cos \theta _{1}\cos \theta _{2}\cos \theta _{3}\cos \theta
_{4}|1^{q_{1}},1^{q_{2}},1^{q_{3}},1^{q_{4}}\rangle  \notag \\
&\otimes &|c^{(0)}\rangle _{A}|c^{(0)}\rangle _{B}|c^{(0)}\rangle
_{C}|c^{(0)}\rangle _{D}  \notag \\
&-&\cos \theta _{1}\cos \theta _{2}\cos \theta _{3}\sin \theta
_{4}|1^{q_{1}},1^{q_{2}},1^{q_{3}},0^{q_{4}}\rangle  \notag \\
&\otimes &|c^{(0)}\rangle _{A}|c^{(0)}\rangle _{B}|c^{(0)}\rangle
_{C}|c^{(1)}\rangle _{D}+\cdots  \notag \\
&+&\sin \theta _{1}\sin \theta _{2}\sin \theta _{3}\sin \theta
_{4}|0^{q_{1}},0^{q_{2}},0^{q_{3}},0^{q_{4}}\rangle  \notag \\
&\otimes &|c^{(1)}\rangle _{A}|c^{(1)}\rangle _{B}|c^{(1)}\rangle
_{C}|c^{(1)}\rangle _{D}.  \label{totalstate}
\end{eqnarray}%
When the control field $E_{c}$ is adiabatically switched off ($\theta
_{i}=\pi /2$, $i=1,2,3,4$), the state of the photonic component of the four
pulses are homogeneously coherently mapped onto the collectively atomic
excitations
\begin{equation}
|1^{q_{1}}\rangle _{1}|1^{q_{2}}\rangle _{2}|1^{q_{3}}\rangle
_{3}|1^{q_{4}}\rangle _{4}\longrightarrow |c^{(1)}\rangle
_{A}|c^{(1)}\rangle _{B}|c^{(1)}\rangle _{C}|c^{(1)}\rangle _{D}.
\label{mapping}
\end{equation}%
According to the polarizations of the input four single photons, from Eq.~(%
\ref{mapping}) we have the following one-to-one mappings:
\begin{equation}
\left(
\begin{array}{c}
|H\rangle _{1}\longrightarrow |c_{-}^{(1)}\rangle _{A} \\
|H\rangle _{2}\longrightarrow |c_{-}^{(1)}\rangle _{B} \\
|H\rangle _{3}\longrightarrow |c_{-}^{(1)}\rangle _{C} \\
|H\rangle _{4}\longrightarrow |c_{-}^{(1)}\rangle _{D}%
\end{array}%
\right) ,\ \ \ \left(
\begin{array}{c}
|V\rangle _{1}\longrightarrow |c_{+}^{(1)}\rangle _{A} \\
|V\rangle _{2}\longrightarrow |c_{+}^{(1)}\rangle _{B} \\
|V\rangle _{3}\longrightarrow |c_{+}^{(1)}\rangle _{C} \\
|V\rangle _{4}\longrightarrow |c_{+}^{(1)}\rangle _{D}%
\end{array}%
\right) .
\end{equation}%
Hence, the state $|\psi \rangle _{\text{atom}}$\ of four atomic
sub-ensembles ($A$, $B$, $C$ and $D$) will depend on the polarization of the
input photons. When the input state is a polarization-encoded cluster state,
after adiabatically turning off the control field, the state of the four
atomic ensembles is a cluster-type state:
\begin{eqnarray}
|\psi \rangle _{\text{atom}} &=&\frac{1}{2}[|c_{-}^{(1)}\rangle
_{A}|c_{-}^{(1)}\rangle _{B}|c_{-}^{(1)}\rangle _{C}|c_{-}^{(1)}\rangle
_{D}+|c_{+}^{(1)}\rangle _{A}|c_{+}^{(1)}\rangle _{B}  \notag \\
&&\otimes |c_{-}^{(1)}\rangle _{C}|c_{-}^{(1)}\rangle
_{D}+|c_{-}^{(1)}\rangle _{A}|c_{-}^{(1)}\rangle _{B}|c_{+}^{(1)}\rangle
_{C}|c_{+}^{(1)}\rangle _{D}  \notag \\
&&-|c_{+}^{(1)}\rangle _{A}|c_{+}^{(1)}\rangle _{B}|c_{+}^{(1)}\rangle
_{C}|c_{+}^{(1)}\rangle _{D}].
\end{eqnarray}%
That is to say that the entangled photon state $|\phi
_{\text{in}}\rangle $ is
coherently mapped to the entangled atomic state $|\psi \rangle _{\text{atom}%
} $. At a later time, the entangled photon state can be retrieved on demand
from the entangled atomic state by turning on $E_{c}$ ($\theta _{i}=0$, $%
i=1,2,3,4$). After passing through the $\lambda /4$ plates again,
the retrieval polarization-encoded cluster state is
\begin{eqnarray}
|\phi _{\text{out}}\rangle &=&\frac{1}{2}[|H\rangle _{1}|H\rangle
_{2}|H\rangle _{3}|H\rangle _{4}+|V\rangle _{1}|V\rangle _{2}|H\rangle
_{3}|H\rangle _{4}  \notag \\
&+&|H\rangle _{1}|H\rangle _{2}|V\rangle _{3}|V\rangle _{4}-|V\rangle
_{1}|V\rangle _{2}|V\rangle _{3}|V\rangle _{4}].~~~~
\end{eqnarray}%
We describe an ideal transfer of a polarization encoded cluster
between light fields and metastable states of atoms \cite{Twocloud}.
In the ideal case, the retrieved pulses are identical to the input
pulses, provided that the same control power is used at the storage
and the retrieval stages. However, to realize the ideal storage, two
conditions must be met: (1) the whole pulse must spatially
compressed into the atomic ensemble, and (2) all spectral components
of the pulse must fit inside the EIT transparency window. In
Secs.~\ref{analysis}, \ref{lifetime} and~\ref{Dis}, we consider the
realistic parameters of the proposal of realization.

Next, we describe the measurement and verification processes of the
retrieval cluster state. Recently, Enk \emph{et al.} \cite{Enk}
discussed a number of different entanglement-verification protocols:
teleportation, violation Bell-Clauser-Horne-Shimony-Holt (CHSH)
inequalities, quantum state tomography, entanglement witnesses
\cite{Toth}, and direct measurements of entanglement. As for the
four-qubit cluster state, the entanglement is verified by measuring
the entanglement witness $\mathcal{W}$. The expectation value of
$\mathcal{W}$ is positive for any separable state, whereas its
negative value detects four-party entanglement close to the cluster
state. The theoretically
optimal expectation value of $\mathcal{W}$ is Tr($\mathcal{W}\rho _{\text{%
theory}}$)$=-1$ for the cluster state \cite{Toth}.

\section{Analysis of Efficiency and Fidelity}

\label{analysis} In this section, we analyze the efficiency and the fidelity
of the memory. The memory efficiency is defined as the ratio of the number
of retrieved photons to the number of incident photons \cite%
{Gorshkov07PRL,Gorshkov07PRA,Gorshkov08PRA,Phillips}:
\begin{equation}
\eta =\int_{T+\tau }^{\infty }|\hat{\mathcal{E}}_{\text{out}%
}(t)|^{2}dt/\int_{0}^{T}|\hat{\mathcal{E}}_{\text{in}}(t)|^{2}dt,
\end{equation}%
where $\tau $ is a storage period. Recently, several proposals are presented
\cite{Gorshkov07PRL,Gorshkov07PRA,Gorshkov08PRA} for the optimal efficiency
of light storage and retrieval under the restrictions and limitations of a
given system. Based on these proposals, two optimal protocols have been
demonstrated experimentally \cite{Novikova07,Novikova08}. The first protocol
iteratively optimizes the input pulse shape for any given control field \cite%
{Novikova07}, while the second protocol uses optimal control fields
calculated for any given input pulse shape \cite{Novikova08}. As for our
cluster storage situation, it is difficult to shape the input signal pulses.
Then the second protocol \cite{Novikova08,Phillips} could be used to improve
the efficiency of storage and retrieval of given signal pulses.

Using the method introduced by Gorshkov \emph{et al.}
\cite{Gorshkov07PRA}, we plot the optimal efficiency $\eta $ of any
one field of the four input fields due to the four fields are
theoretically equivalent. When considering the spin wave decay, one
should just multiply the efficiency by $\exp (-2\gamma _{s}\tau )$
\cite{Gorshkov07PRA} and the efficiency will decrease.
Figure~\ref{efficiency} shows the optimal efficiency $\eta $ versus
the optical depth $d$ under different spin wave decays $\gamma
_{s}$. It is desirable to read out as fast as possible due to the
spin wave decay ($\tau \ll 1/\gamma _{s}$).

Next, we analyze the fidelity of memory. For the density matrices $\rho
_{in} $ and $\rho _{out}$ of the input and output quantum state, the
fidelity is defined as \cite{Uhlmann,NielsenBook}:
\begin{equation}
F=(\mathtt{Tr}\sqrt{\rho _{\text{in}}^{1/2}\rho _{\text{out}}\rho _{\text{in}%
}^{1/2}})^{2}.
\end{equation}%
For two pure states, their fidelity coincides with the overlap. This number
is equal to one only for the ideal channel transmitting or storing the
quantum state perfectly. We describe a method of the projector-based
entanglement witness (Tr($\mathcal{W}\rho _{\text{exp}}$)) in Sec.~\ref%
{polarization} to verify the cluster state, which is also used to obtain
information about the fidelity $F=\langle \phi _{\text{in}}|\rho _{\text{exp}%
}|\phi _{\text{in}}\rangle $ \cite{Toth,Kiesel,Vallone} from the
measurement process. The observed fidelity $F>1/2$ assures that the
retrieved state has genuine four-qubit entanglement
\cite{Toth,Tokunaga}. The high-fidelity quantum memory is necessary
for quantum information and quantum communication. Due to
nonsimultaneous retrieval, the influence on the fidelity will be
discussed in Sec. \ref{Dis}.

\section{Lifetime of quantum memory}

\label{lifetime}

\begin{figure}[tbp]
\centerline{\includegraphics[scale=0.6,angle=0]{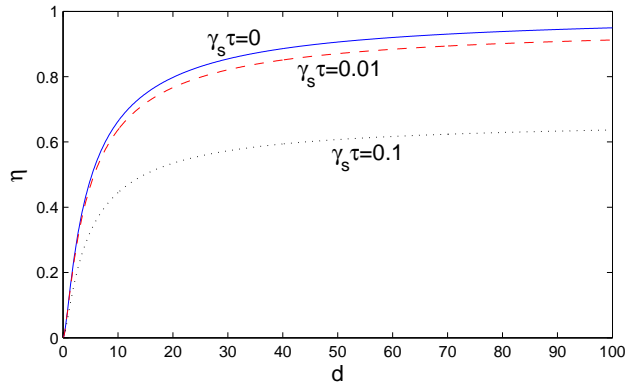}}
\caption{ (Color online) $\protect\eta$ is the optimal efficiency for
adiabatic storage and retrieval with dimensionless spin wave decay $\protect%
\gamma_{s}\protect\tau$.}
\label{efficiency}
\end{figure}

In this section, we discuss the lifetime of the proposed memory.
Quantum memories for storage and retrieval of quantum information
are extremely sensitive to environmental influences, which limits
their storage time. In many experiments of cold atoms in the MOT,
the quadrupole magnetic field is
the main source of the atomic memory decoherence \cite{Matsukevich,Polyakov}%
. Using the freely cold rubidium atoms released from MOT \cite%
{MatsukevichPRL06}, the coherence time will be increased \cite%
{MatsukevichPRL05} and the longest quantum memory time of the system
reported to date is $32$ $\mu $s \cite{MatsukevichPRL06} without
using the clock states, where the time is limited by dephasing of
different Zeeman components in the residual magnetic field.\
Recently, using magnetic field insensitive states (clock states),
the storage time of the quantum memory
storing single excitation has improved to $1$ ms \cite{Zhaobo}. Zhao \textit{%
et al.} \cite{Zhao} also used the magnetically insensitive clock transition
for the quantum memory, but they confined the cold rubidium atomic ensemble
in a one-dimensional optical lattice, and the memory lifetime has improved
and exceeded $6$ ms.

In our scheme, all the light fields responsible for trapping and
cooling, as well as the quadrupole magnetic field in the MOT, are
shut off during the period of the storage and retrieval process;
ballistic expansion of the freely falling gas provides a longer
memory time limitation. Assuming
one-dimensional case and Maxwell-Boltzmann velocity distribution $f(v_{z})=%
\sqrt{M/2\pi k_{B}T}e^{-mv_{z}^{2}/2k_{B}T}$, the atomic mean speed is $%
\langle v\rangle =\sqrt{k_{B}T/M}$ where $k_{B}$ is the Boltzmann
constant, $T$ is the temperature, and $M$ is atomic mass. After a
storage period $\tau $, the collective state describing the spin
wave $|c^{(1)}(t)\rangle $ evolves to
\begin{equation}
|c^{(1)}(t+\tau )\rangle =\frac{1}{\sqrt{N}}\sum_{j=1}^{N}e^{i\Delta
k(z_{j}+v_{j}\tau )}|b_{1},b_{2},...,c_{j},...,b_{N}\rangle ,
\end{equation}%
where $\Delta k=k_{p}-k_{c}^{||}+q$ the wavevector of the spin wave. For
narrow transparency window $\Delta \omega _{tr}$, $\left\vert q\right\vert
\ll k$, the influence of field mode $q$ on the storage time can be ignored.
The stored information due to atomic random motion begins to decrease, and
the obtainable information by retrieval process is proportional to
\begin{eqnarray}
R&=&\left\vert \langle c^{(1)}(t)|c^{(1)}(t+\tau )\rangle \right\vert
^{2}=\left\vert \frac{1}{N}\sum_{j=1}^{N}e^{i\Delta kv_{j}\tau }\right\vert
^{2}  \notag \\
&=&\left\vert \int f(v)e^{i\Delta kv\tau }dv\right\vert ^{2}\approx \exp (-%
\frac{\tau ^{2}}{\tau _{s}^{2}}),
\end{eqnarray}%
where the integration limits $\pm \infty $ are used to obtain the analytic
solution and $\tau _{s}$ is the $e^{-1}$-coherence time given as \cite%
{Yoshikawa05}
\begin{equation}
\tau _{s}=\frac{1}{\Delta k\langle v\rangle }.  \label{SPlifetime}
\end{equation}%
The dephasing induced by atomic random motion can also be described by the
grating wavelength of the spin wave \cite{Zhaobo,Zhao}
\begin{equation}
\lambda _{s}=\frac{2\pi }{\Delta k}.  \label{spinwavelength}
\end{equation}%
The parameters of a suitably cold atomic cloud provided by a MOT \cite%
{MatsukevichPRL05,MatsukevichPRL05} for our proposal are as follows:
the optical depth of about $d=10$, the temperature of $T=70$ $\mu
$K, and the atomic mean speed $\langle v\rangle
=\sqrt{k_{B}T/M}\simeq 8$ cms$^{-1}$ for rubidium mass $M$. From
Eq.~(\ref{SPlifetime}), the lifetime of spin wave can be tuned by
varying the wave vector of the spin wave.

In case $1$ where the control field and the probe fields are
propagating in orthogonal direction, $\Delta k\simeq k_{p},$ the
spin wave grating wavelength is about light wavelength and the spin
wave would dephase rapidly due to atomic motion, so $\tau _{s}\simeq
1.5$ $\mu $s. Considering the efficiency of memory, the storage time
is less than 1 $\mu $s.

In case $2$ the Doppler-free configuration propagation, almost no
photonic momentum is transferred to the atoms. The atomic coherence
is localized in the longitudinal direction, in which oscillations
have the small beat frequency $\Delta \omega \simeq
(k_{p}-k_{c}^{||})c=6.8$ GHz between the probe and the control
fields, and the calculated spin-wave wavelength by
Eq.~(\ref{spinwavelength}) is $\lambda _{s}\approx 27$ cm. The
dephasing induced by the atomic motion in this localized region is
very slowly due to
the large spin-wave wavelength $\lambda _{s}$, and the computed lifetime $%
\tau _{s}$\ is large. However atomic random motion would spread the
localized excitation from one ensemble to another ensemble, which
will result in the stored information being quickly lost.
Considering the atoms flying out of the localized atomic ensemble
and the waist of laser beams $D=100$
$\mu $m, the lifetime of the memory can be estimated as $\tau =D/(2v)\sim 300$ $%
\mu $s.

Another factor that influences the memory time is the decoherence of
the excited state. The DSP is protected against incoherent decay
processes acting on the excited states because of adiabatical
elimination of the excited states. Although the collective state
$|c^{(1)}\rangle $ is an entangled state of $N$ atoms, its
decoherence time is not much different from that of the quantum
state stored in an individual atom and it is quite stable against
one-atom (or few-atom) losses \cite{Mewes}. Due to the memory
efficiency, the maximum storage time of our proposal must be far smaller than $%
\tau _{s}$ ($\tau \ll 1/\gamma _{s}$), or else the efficiency will
be low. From Fig.~\ref{efficiency}, in order to balance the
efficiency and preserve the entanglement, suitable storage times for
cases $1$ and $2$ are less than or equal to $0.15$ and $30$ $\mu $s,
respectively.

\section{Discussion}

\label{Dis} In this section, we discuss some restrictions. First, we
assume that the input the four-photon cluster state $|\phi
_{\mathtt{in}}\rangle $
generated by experiment \cite%
{clusterexp,Kiesel,Zhang,Prevedel,Lu,Vallone,Chenprl,Tokunaga} has
high fidelity. Then the four-photon cluster state should be sent
into the atomic cloud simultaneously, which requires that the four
incidence points are symmetric about major axis of ellipsoid. If
there is a large difference, one single-photon probe field is not
synchronously sent into the atomic ensemble with other fields. Then
a fraction of single-photon wave packet, captured in the form of a
spin wave, is stored for a time period $\tau $. The efficiency of
light storage will decrease, and the shape of the output pulse is
different from the initial pulse. This case can be avoided using the
perfect cluster state and choosing symmetric atomic ensembles.
Second, in the retrieval process, we assume that one stored field is
not simultaneously retrieved or is not retrieved even. If the four
fields are retrieved non-simultaneously and only one field is
retrieved with a little delay, the entanglement is preserved and the
entanglement degree decreases \cite{Curtis}. If one field is
retrieved with
a certain probability, without loss of generality, we also choose field $%
E_{1}$. After the retrieval, the field $E_{1}$ can be written as
$(|0\rangle _{1}+\beta _{1}|1\rangle _{1})/\sqrt{1+|\beta
_{1}|^{2}}$, so the fidelity is $F=|\langle \phi _{\text{in}}|\phi
_{\text{out}}^{\prime }\rangle |^{2}=|\beta _{1}|^{2}/(1+|\beta
_{1}|^{2})$. The retrieval state is a cluster state provided that
$|\beta _{1}|^{2}>1$ \cite{Toth,Tokunaga}. In order to retrieve the
frequency-entangled state with high fidelity $F>95\%$, then the
coefficient $|\beta _{1}|^{2}$ should be necessarily more than $20$.
That is to say that the four fields would be retrieved nearly
simultaneously, or else the fidelity is low. The third factor that
limits the performance of the storage is adiabatic condition.
Adiabatic following occurs when the population in the excited and
bright states is small at all times. For a pulse duration $T$ and a
line-width of the excited state $\gamma $, the adiabatic condition
is $g^{2}N\gg \gamma /T$.

\section{Conclusion}

In conclusion, we present a scheme for realizing quantum memory for
the four-photon polarization encoded cluster state. Our proposal can
be realized
by current technology \cite%
{Choi,Simon,MatsukevichPRL06,MatsukevichPRL062,Chen07}. The quantum memory
of the cluster is essential for \textquotedblleft one-way\textquotedblright\
quantum computing, and we also expect the ability to store multiple optical
modes to be useful for the quantum information and all-optical quantum
computing network.



\begin{thebibliography}{99}
\bibitem{Twocloud} This cluster state memory can also be realized by mapping
the four pulses in and out by extending one cold cloud to two cold clouds in
order to avoid the light fields spatial overlapping, each cloud mapping two
pulses.
\end{thebibliography}
\end{document}